%% file: main.tex
\documentclass{optica-article}

\journal{Optica}

\articletype{Research Article}

\usepackage{hyperref}
\hypersetup{colorlinks,
	linkcolor={blue!75!black!80!yellow},
	citecolor={blue!75!black!80!yellow},
	urlcolor={blue!75!black!80!yellow}
}

\usepackage{xcolor}
\definecolor{orange}{RGB}{217,83,25}
\definecolor{blue}{RGB}{0,114,189}
\definecolor{magenta}{RGB}{200,0,200}
\definecolor{green}{RGB}{0,146,69}

\input{misc-utils.tex}
\setboolean{togglecomments}{true} 
\setboolean{toggletodos}{true}
\setboolean{togglechanges}{false} 

\usepackage{lineno}

\begin{document}

\title{Topological electromagnetic waves in dispersive and lossy plasma crystals}

\author{Chen~Qian,\authormark{1,5} Yue~Jiang,\authormark{1,5} Jicheng~Jin,\authormark{1} Thomas~Christensen,\authormark{2,4} Marin~Solja\v{c}i\'c,\authormark{2} Alexander~V.~Kildishev,\authormark{3}and Bo~Zhen\authormark{1,*}}

\address{\authormark{1} Department of Physics and Astronomy, University of Pennsylvania, Philadelphia, Pennsylvania, 19104, USA \\
\authormark{2} Department of Physics, Massachusetts Institute of Technology, Cambridge, Massachusetts, 02139, USA\\
\authormark{3} Birck Nanotechnology Center and Purdue Quantum Science and Engineering Institute (PQSEI), Elmore Family School of Electrical and Computer Engineering, Purdue University, West Lafayette, Indiana, 47907, USA\\
\authormark{4}Department of Electrical and Photonics Engineering, Technical University of Denmark, 2800 Kgs.~Lyngby, Denmark\\
\authormark{5}The authors contributed equally to this work.\\}

\email{\authormark{*}bozhen@sas.upenn.edu} 



\begin{abstract}
Topological photonic crystals, which offer topologically protected and back-scattering-immune transport channels, have recently gained significant attention for both scientific and practical reasons. Although most current studies focus on dielectric materials with weak dispersions, this study focuses on topological phases in dispersive materials and presents a numerical study of Chern insulators in gaseous-phase plasma cylinder cells. We develop a numerical framework to address the complex material dispersion arising from the plasma medium and external magnetic fields and identify Chern insulator phases that are experimentally achievable. Using this numerical tool, we also explain the flat bands commonly observed in periodic plasmonic structures, via local resonances, and how edge states change as the edge termination is periodically modified. This work opens up opportunities for exploring band topology in new materials with non-trivial dispersions and has potential RF applications, ranging from plasma-based lighting to plasma propulsion engines.
\end{abstract}

\section{Introduction}
The discovery that topological phases can exist beyond electronics~\cite{haldane2008possible, wang2009observation, JOUR2} has led to significant interest in other wave systems such as photonics~\cite{zero-frequency_2, JOUR, PhysRevLett.126.113901}, plasmonics~\cite{jin2016topological, jin2017infrared, pan2017topologically}, polaritonics~\cite{li2022polaritonic}, acoustics~\cite{yang2015topological, JOUR3}, and even water waves~\cite{delplace2017topological, perrot2019topological}. Of particular interest is the quantum anomalous Hall effect, also known as Chern insulators, which can protect unidirectional and back-scattering-immune transport channels at the interface with normal insulators. Such transport channels have both fundamental and practical applications in fields such as optical communication, low-loss waveguides, and circulators.

To create photonic Chern insulators, time-reversal symmetry must be broken while retaining near-Hermiticity. This requires a permittivity or permeability tensor that breaks reciprocity, such that $\boldsymbol{\varepsilon}^{\text{T}}\ne \boldsymbol{\varepsilon}$ or $\boldsymbol{\mu}^{\text{T}}\ne \boldsymbol{\mu}$. The properties of the underlying materials, along with their geometric aspects, determine the topological invariants (Chern numbers) of the electromagnetic band gaps and their transport properties.

Most current studies of photonic Chern insulators have relied on gyromagnetic materials (e.g. yttrium iron garnet~\cite{wang2009observation, PhysRevLett.100.013905, PhysRevLett.113.113904, liu2022topological}) and external magnetic fields. These materials are mostly dielectric in nature, meaning their permeability remains positive in the frequency range of interest, resulting in Bloch modes that are delocalized in the photonic crystals. There has been recent interest in gyroelectric materials, which have the potential for large Faraday effects in magnetized plasmas in metals. This focus has centered on continua~\cite{PhysRevB.94.205105} and plasmonic crystals~\cite{jin2016topological, jin2017infrared, pan2017topologically}, where the Chern insulator phase is mostly composed of coupled plasmonic resonances localized at individual sites with Drude-like material responses and negative permittivity.

Here, we present a numerical study of photonic Chern insulators in plasma crystals with gaseous-phase plasma, which simultaneously exhibit both extended photonic bands and localized plasmonic modes in the RF regime.  
Our design is based on a 2D crystal of plasma cells placed in an external magnetic field. 
Without the magnetic field, the responses of the plasma elements are Drude-like, and the associated structure is known to support coexisting de-localized and localized modes in suitable polarizations~\cite{sakoda2001photonic-I, sakoda2001photonic-II, PhysRevB.83.205131}. 
We explore the time-reversal broken generalization, which exhibits an interesting interplay between the Drude and Lorentzian dispersion due to the applied magnetic field causing cyclotron motions in the plasma. 
We propose a plasma crystal design that features a Chern insulator gap between de-localized photonic bands, coexisting with nearby dense groups of flat bands associated with localized plasmons of fixed-handedness. 
On termination of this crystal, we observe a rich interplay between localized plasmonic bands and de-localized chiral edge states inside the Chern insulating gap. 
Finally, we explore how local and de-localized edge modes evolve under continuous deformations of the interface between the Chern insulator and perfect magnetic conductors. The mode evolution can then be interpreted as a manifestation of the filling anomalies. 

Our work is organized as follows: in Sections~\ref{sec:drude-zero-B} and \ref{sec:drude-nonzero-B}, we review the plasma dispersion without and with an external magnetic field. 
In Sections~\ref{sec:bandstructure-zero-B} and \ref{sec:bandstructure-nonzero-B}, we present the band structure of plasma crystals without and with an external magnetic field and their associated topological invariants. 
In Section~\ref{sec:flat-bands}, we explain the origin of the observed flat bands in calculations as localized surface plasmon polariton resonances. 
In Section~\ref{sec:evolution}, we explore how the chiral edge state dispersion evolves when the edge termination changes. 
Finally, we discuss the limitations existing in our calculations and practical aspects related to the experimental verification of our proposal. 

\section{Dispersion of plasma without magnetic field: Drude model}
\label{sec:drude-zero-B}

We start by reviewing the dispersion of the plasma permittivity $\varepsilon$, following the standard Drude model~\cite{Swanson}. 
Assuming that the positive ions are too heavy to move, the volume current density is solely contributed by electrons: $\textbf{J} = -n e \textbf{v}$. Here $n$ is the volume density of electrons, and $-e$ is the electron charge. 
The equation of motion for electrons in the plasma crystals reads:  
\begin{equation}
    m_{\rm e} \partial_t \textbf{J} +  \gamma m_{\rm e} \textbf{J} = ne^{2} \textbf{E}. 
\end{equation}

Here, $m_{\rm e}$ is the electron mass, $\gamma$ is the damping rate, $n$ is the number density of electrons, and $\textbf{E}$ is the electric field. 
For harmonic solutions at a fixed angular frequency $\omega$, all temporal derivatives can be substituted via $\partial_t \rightarrow -\mathrm{i}\omega$. 
Accordingly, the frequency-dependent plasma conductivity $\sigma$ can be written as
\begin{equation}
\textbf{J} = \left( \frac{\varepsilon_0 \omega^2_{\rm p}}{-\mathrm{i}\omega + \gamma} \right) \textbf{E} = \sigma \textbf{E}. 
\end{equation} 
Here, $\omega_{\rm p} = \sqrt{\frac{ne^2}{m_{\rm e}\varepsilon_0}}$ is the plasma frequency. 
Noting that the volume current density is also related to the electric polarization: $\textbf{J} = \partial_t \textbf{P} = -\mathrm{i}\omega \textbf{P}$, the Drude permittivity of plasma $\varepsilon$ can be defined as:
\begin{equation}
   \textbf{D} = \varepsilon_0 \textbf{E} + \textbf{P} = \left( \varepsilon_0 + \frac{\mathrm{i}\sigma}{\omega} \right){\textbf{E}}  = \varepsilon_0 \left[1- \frac{\omega_{\rm p}^2}{\omega(\omega+\mathrm{i} \gamma)} \right]\textbf{E}. 
   \label{eq:Dfield_constitutive}
\end{equation}

\section{Dispersion of magnetized plasma}
\label{sec:drude-nonzero-B}

\begin{figure}[htbp]
    \centering\includegraphics[width=\linewidth]{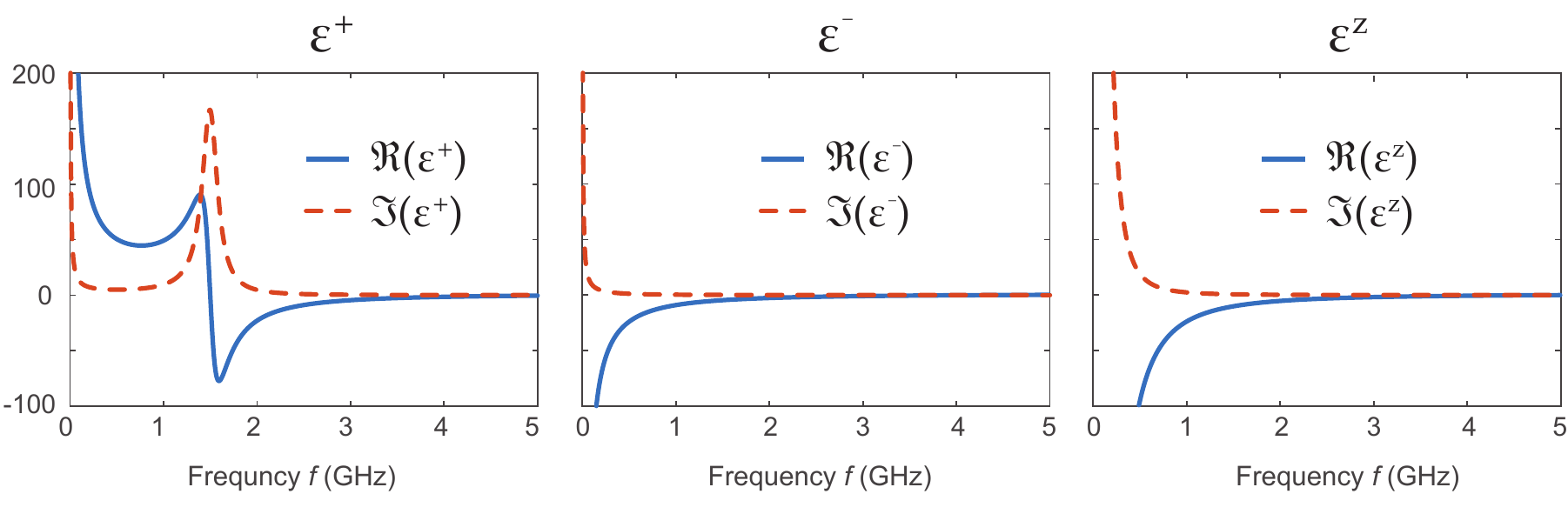}
    \caption{%
        \textbf{Permittivity dispersion of plasma medium in an external magnetic field. }
        $\varepsilon_+$ ($\varepsilon_-$) is the permittivity of right-handed (left-) circular polarization under an external magnetic field of 0.054\,T.
        $\varepsilon_z$ is for polarization along the $z$ direction. 
        The real and imaginary parts of the permittivity are shown in blue and red, respectively. 
        Both $\varepsilon_{+}$ and  $\varepsilon_{-}$ are affected by the cyclotron resonance at $f_{\rm c} = 1.5\,\text{GHz}$, while $\varepsilon_z$ is unaffected and remains to be given by the standard Drude model.
        \label{fig:permittivity}
        }
\end{figure}

Following similar steps, we next show the permittivity tensor describing gaseous phase plasma placed in an external magnetic field (Fig.~\ref{fig:permittivity}a).
Following Eq.~\eqref{eq:Dfield_constitutive}, we need to re-write the conductivity tensor ($\Bar{\Bar{\sigma}}$), based on the updated equation of motion for electrons.  

Taking into account the Lorentz force, the equation of motion becomes: 
\begin{equation}
    m_{\rm e} \partial_t\textbf{J} + m_{\rm e} \gamma \textbf{J} = ne^2 \textbf{E} + e\textbf{J}\times B\hat{\mathbf{z}}
\end{equation}
Re-writing the equation using circular bases in the $xy$ plane:  $(J_+, J_-, J_z) = (\frac{J_x+\mathrm{i}J_y}{\sqrt{2}}, \frac{J_x-\mathrm{i}J_y}{\sqrt{2}}, J_z)$ and $(E_+, E_-, E_z) = (\frac{E_x+\mathrm{i}E_y}{\sqrt{2}}, \frac{E_x-\mathrm{i}E_y}{\sqrt{2}}, E_z)$, both matrices $\Bar{\Bar{\sigma}}$ and $\Bar{\Bar{\varepsilon}}$ become diagonal. Specifically, the conductivity tensor $\Bar{\Bar{\sigma}}$ can be expressed as: 
\begin{equation}\label{eq:current_ewfd}
\begin{pmatrix}
J_{+} \\
J_{-} \\
J_{z} 
\end{pmatrix} 
= \mathrm{i}\varepsilon_0 \omega_{\rm p}^2
\begin{pmatrix}
\frac{1}{\omega + \mathrm{i} \gamma - \omega_{\rm c}} &  & \\
 & \frac{1}{\omega + \mathrm{i} \gamma + \omega_{\rm c}} & \\
 &   & \frac{1}{\omega + \mathrm{i} \gamma} 
\end{pmatrix}
\begin{pmatrix}
E_{+} \\
E_{-} \\
E_{z}
\end{pmatrix}.
\end{equation}
Here $\omega_{\rm c} = \frac{eB}{m_{\rm e}}$ is the cyclotron resonance.
Accordingly, the permittivity tensor of the magnetized plasma $\Bar{\Bar{\varepsilon}}$ has the following form in a Cartesian basis: 
\begin{equation}
\bar{\bar{\varepsilon}} = \textbf{U}^{\dagger}
\begin{pmatrix}
\varepsilon_{+} &  & \\
 & \varepsilon_{-} & \\
 &   & \varepsilon_{z} 
\end{pmatrix}
\textbf{U}, 
\end{equation}
where the column vectors of the matrix $U$ label the directions of the optical principle axes: 
\begin{equation}
\textbf{U} =  
\begin{pmatrix}
\frac{1}{\sqrt{2}} & \frac{\mathrm{i}}{\sqrt{2}} & \\
\frac{\mathrm{i}}{\sqrt{2}} & \frac{1}{\sqrt{2}} & \\
 &   & 1
\end{pmatrix}. 
\end{equation}
Meanwhile, along the principal axes, the material dispersion can be expressed as: 
\begin{equation} \label{magetized-dispersion}
\begin{split}
 \varepsilon_{\pm} & = \varepsilon_0 + \mathrm{i} \frac{\sigma_\pm}{\omega} = \varepsilon_0 \left[1 - \frac{\omega_{\mathrm{p}}^2 }{\omega (\omega  + \mathrm{i} \gamma \mp \omega_{\mathrm{c}})} \right], \\
 \varepsilon_{z}  & = \varepsilon_0 + \mathrm{i} \frac{\sigma_z}{\omega} = \varepsilon_0 \left[1 - \frac{\omega_{\mathrm{p}}^2}{\omega(\omega + \mathrm{i} \gamma)} \right]. 
\end{split}
\end{equation}
 This result can be intuitively understood as follows: under the influence of the external magnetic field along $\hat{\mathbf{z}}$, electrons undergo cyclotron motion and orbit at a fixed angular frequency of $\omega_{\mathrm{c}}$ in the $xy$ plane. In the frame co-rotating (counter-rotating) with the electrons, the driving electric field of the same (opposite) circular polarization is thus offset in frequency by $\omega_{\rm c}$ ($-\omega_{\rm c}$), giving rise to $\varepsilon_+$ ($\varepsilon_-$). Meanwhile, the applied electric field in the $z$ direction is not affected by the cyclotron motion, so $\varepsilon_z$ remains the standard Drude dispersion.  

Next, we numerically compute the material dispersion as functions of frequency $f = \omega/2\pi$ using the typical values in gaseous phase plasma. The results are shown in Fig.~\ref{fig:permittivity}, where the external magnetic field is set at 0.054\,T and the corresponding cyclotron resonance is at $f_{\rm c} = \omega_{\rm c}/2\pi = 1.5\,\text{GHz}$. The plasma frequency is controlled by the number density of electrons, which is set to be a practical value of $n=2.8\times10^{11}\,\text{cm}^{-3}$ throughout the calculations. 
Accordingly, 
the plasma frequency is at $f_{\rm p} = \omega_{\rm p}/2\pi = 5\,\text{GHz}$.
The real and imaginary parts of the permittivity are shown in blue and red, respectively. The parameter $\gamma$ refers to the damping rate, which mostly originates from the electron-ion collisions at room temperature. For helium plasma gas, the damping rate is roughly linearly proportional to the gas pressure: $\gamma = 2\,\text{GHz}\times p [\text{Torr}]$. Using a practical pressure of $p= 0.05\,\text{Torr}$, the damping is set to be $\gamma = 0.1\,\text{GHz}$ throughout our calculations.

As expected from Eq.~\eqref{magetized-dispersion}, both 
$\varepsilon_+$ and $\varepsilon_-$ are affected by the cyclotron resonance and deviate from the standard Drude model ($\varepsilon_z$). For example, at low frequencies ($f \approx 0$), $\varepsilon_+$ is positive and diverges as $1/f$; $\varepsilon_{-}$ is negative and diverges as $1/f$; meanwhile, the Drude model $\varepsilon_z$ is negative and diverges as $1/f^2$. These fundamental differences in scaling lead to challenges when fitting the dispersion to standard formalism in commercial software, as described later. 

We note that our description of the plasma medium is limited by a few approximations. Overcoming these approximations would lead to modifications of our results and will be discussed elsewhere. First, only electrons are assumed to move under the influence of external electromagnetic fields, while ions are assumed to be always stationary. Second, our permittivity ignores non-local effects, which leads to a frequency gap between surface plasmon polaritons traveling in opposite directions\cite{PhysRevB.55.7427, HassaniGangaraj:19, buddhiraju2020absence}. Finally, we neglect the inhomogeneous broadening effect, and thus the bandwidth of our permittivity is defined solely by electron damping.

\section{Band structures of plasma crystals without external magnetic field} 
\label{sec:bandstructure-zero-B}

Based on the dispersion in Eq.~\eqref{magetized-dispersion}~\cite{Howard}, we compute the band structure of a square lattice of plasma cylinders placed in the air. Specifically, we focus on quadratic point degeneracy, which is protected by spatial symmetry and time-reversal symmetry.  

\begin{figure}[htbp]
    \centering
    \includegraphics[width=0.5\linewidth]{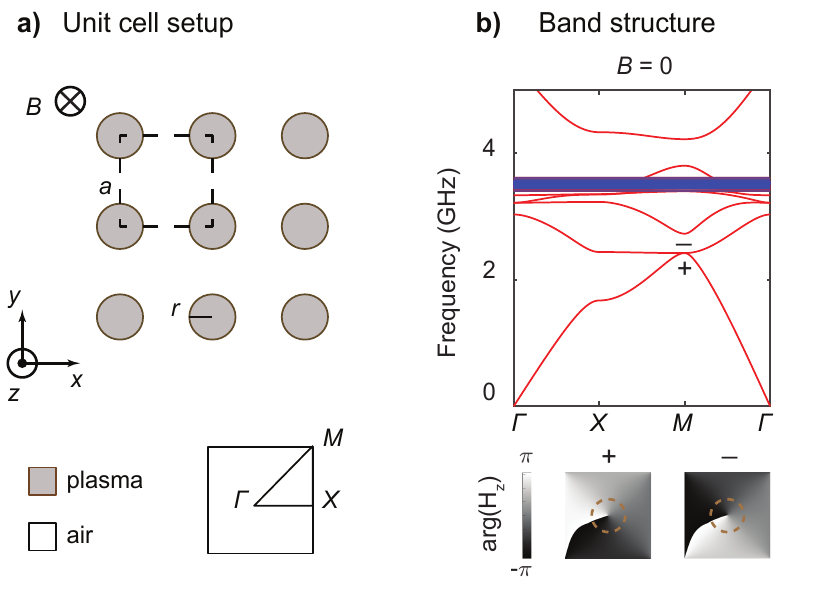}
    \caption{%
        \textbf{Band structure of a plasma photonic crystal without external magnetic field.} 
        \textbf{(a)}~Schematic drawing of a photonic crystal made of gaseous plasma cylinders placed in the air.
        \textbf{(b)}~Calculated H-case without magnetic field ($B=0$), where a quadratic degeneracy, modes `$+$' and `$-$', with $C_{4z}=\pm i$ respectively, is found at the Brillouin zone corner.
        A set of flat bands (blue ribbon) are also observed in the calculation. 
        \label{fig:bandstructure_timereversal}
    }
\end{figure}

The plasma photonic crystal unit cell is shown in Fig.~\ref{fig:bandstructure_timereversal}a, where the lattice constant $a$ is 6\,cm, and the radius of the cylinder $r$ is 1.5\,cm.
The external magnetic field along the $z$ direction preserves the mirror symmetry in $z$ ($\sigma_z$) and separates the electromagnetic modes into two mode types:
to avoid possible terminology confusion, we follow the definition of Sakoda~\cite{Sakoda} to distinguish the E-polarization case (with the $\mathbf{E}$-field parallel to the $z$-axis) from the H-polarization case (with the $\mathbf{H}$-field parallel to the $z$-axis).  We also use the E-case and H-case for short. Given that the gyroelectric response has response components only in the $xy$-plane but not along the $z$-direction [Eq.~\eqref{eq:current_ewfd}], we focus solely on the H-case and disregard the associated E-case.

Without breaking time-reversal symmetry ($T$), i.e., $B = 0$, the plasma dispersion follows the standard Drude model, and the H-case can be calculated using a standard finite element method (FEM). The results are plotted along high-symmetry lines in the Brillouin zone (Fig.~\ref{fig:bandstructure_timereversal}b). As shown, a pair of quadratic degeneracies are found at the $M$ point in the Brillouin zone around 2.4\, GHz, which is protected by the 90-degree rotation symmetry $C_{4z}$ and $T$. Specifically, the two modes, marked as  `$+$' and `$-$', have $C_{4z}$ indices of $\pm i$,. They are connected to each other by $T$. The phases of the corresponding mode profiles, $\mathop{\mathrm{arg}}(H_z)$, confirm the $C_{4z}$ indices of the two modes. We note that a set of flat bands are observed in the calculation (blue ribbon) with an upper-frequency bound  of $f_{\rm p}/\sqrt{2} = 3.5\,\text{GHz}$, which is further discussed in the next section.  

\section{Band structure of the magnetized-plasma photonic crystals and Chern insulators} 
\label{sec:bandstructure-nonzero-B}

When an external magnetic field is applied, the plasma dispersion ($\varepsilon_{+}$ and $\varepsilon_{-}$) deviates from the Drude model, and the band structure proves difficult to be faithfully captured by using the standard material dispersion in various commercial software. Hence, we modify a standard FEM method to trace the desired number of bands, reliably resolving the band crossings. Similar to conventional approaches (see e.g., \cite{PhysRevB.50.16835, PhysRevB.58.7230, Degirmenci:13, lallane2018prb, Xiao:21}), we employ the Floquet periodic boundary conditions to solve a quadratic eigenvalue problem in a square unit cell depicted in Fig.~\ref{fig:bandstructure_timereversal}a; we also discretize the cell to get a weak FEM formulation. We use commercial FEM software (\textsc{COMSOL Multiphysics\textsuperscript{\texttrademark}}) -- a flexible platform for constructing a system of customized coupled equations that employs the Portable Large Scale Eigenvalue Package (\textsc{P\_ARPACK}). \cite{maschhoff1996} \textsc{P\_ARPACK}, a parallel version of the \textsc{ARPACK} software, \cite{lehoucq1998} implements the Implicitly Restarted Arnoldi Method (IRAM) capable of solving large sparse eigenvalue problems for a select number of additionally constrained eigenvalues. \textsc{COMSOL Multiphysics\textsuperscript{\texttrademark}} also provides a convenient user interface for controlling the key tunable parameters of its IRAM-based scalable eigensolver. Advantages of solving customized quadratic eigenproblems with \textsc{P\_ARPACK} in \textsc{COMSOL} have been proven by numerous studies, see e.g., \cite{davanco2007,fietz2011,lallane2018prb}.

In contrast to the known methods largely employing the auxiliary equations for the polarization vector, we couple \textit{the weak-form equations for the current density and the $\mathbf{E}$-field}, achieving a stable performance with flexible tracking of a desired number of bands and unambiguous resolution of the band crossings.           
Section~\ref{sec:drude-nonzero-B} provides a general equation for current density under an external magnetic field.
In the H-case, we directly use the External Current Density Interface with an auxiliary algebraic equation (AE) for in-plane vectors,
\begin{equation}
\mathbf{J}_{xy} = \left( J_x, J_y \right)^{\mathrm{T}},\;
\mathbf{E}_{xy}= \left(E_x,E_y\right)^{\mathrm{T}}    
\end{equation} 
We then employ a normalized current density $\mathbf{j}_{xy} = \mathbf{J}_{xy}/\left(\varepsilon_0\omega_\mathrm{p}^2\right)$ in a normalized Drude model
\begin{equation}
\left(\omega_{\mathrm{c}}^2 - (\omega + \mathrm{i}\gamma)^2 \right)\mathbf{j}_{xy} - 
\mathrm{i}
\begin{pmatrix}
\omega + \mathrm{i}\gamma & -\mathrm{i}\omega_{\mathrm{c}} \\
\mathrm{i}\omega_{\mathrm{c}} & \omega + \mathrm{i}\gamma 
\end{pmatrix} 
\mathbf{E}_{xy} = 0
\label{eq:Drude_normalized}
\end{equation}
The weak form is obtained by integrating the dot product of Eq.~\eqref{eq:Drude_normalized} with the test function $\mathbf{j}_{xy} = \mathop{\mathrm{test}}(\mathbf{j}_{xy})$ over the Drude material domain,
\begin{equation}
   \int_{\Omega}\left( \left(\omega_{\mathrm{c}}^2 - (\omega + \mathrm{i}\gamma)^2 \right)\mathbf{j}_{xy} - 
\mathrm{i}
\begin{pmatrix}
\omega + \mathrm{i}\gamma & -\mathrm{i}\omega_{\mathrm{c}} \\
\mathrm{i}\omega_{\mathrm{c}} & \omega + \mathrm{i}\gamma 
\end{pmatrix} 
\mathbf{E}_{xy}\right)\cdot\mathbf{j}_{xy}\,\mathrm{d}\Omega = 0 
\end{equation}
An auxiliary equation for the current density is introduced to the \textsc{COMSOL} framework through the Weak Contribution interface,
\begin{equation}
F = \mathrm{i}\omega \textit{k}_{\mathrm{p}}^2 \left(\mathbf{j}_{xy}\cdot\mathbf{E}_{xy}\right)
\end{equation}
where $\mathbf{E}_{xy} = \mathop{\mathrm{test}}(\mathbf{E}_{xy})$ is the $\mathbf{E}$-field test function and $k_{\mathrm{p}}=\omega_{\mathrm{p}}/{c}$ is the plasma wave number. 
The approach exhibits good error convergence with accuracy controlled through the meshing density and the FE order. In contrast with \cite{raman2010photonic, davanco2007, fietz2011, parisi2012, lallane2018prb} we completely exclude the polarization vector, reducing the order of the auxiliary equations and improving the numerical accuracy at $\omega\to 0$. The full details on the numerical implementation and verification of the IRAM-based eigensolver are not the main focus of the paper and these details will be published elsewhere.

\begin{figure}[htbp]
    \centering
    \includegraphics[width=\linewidth]{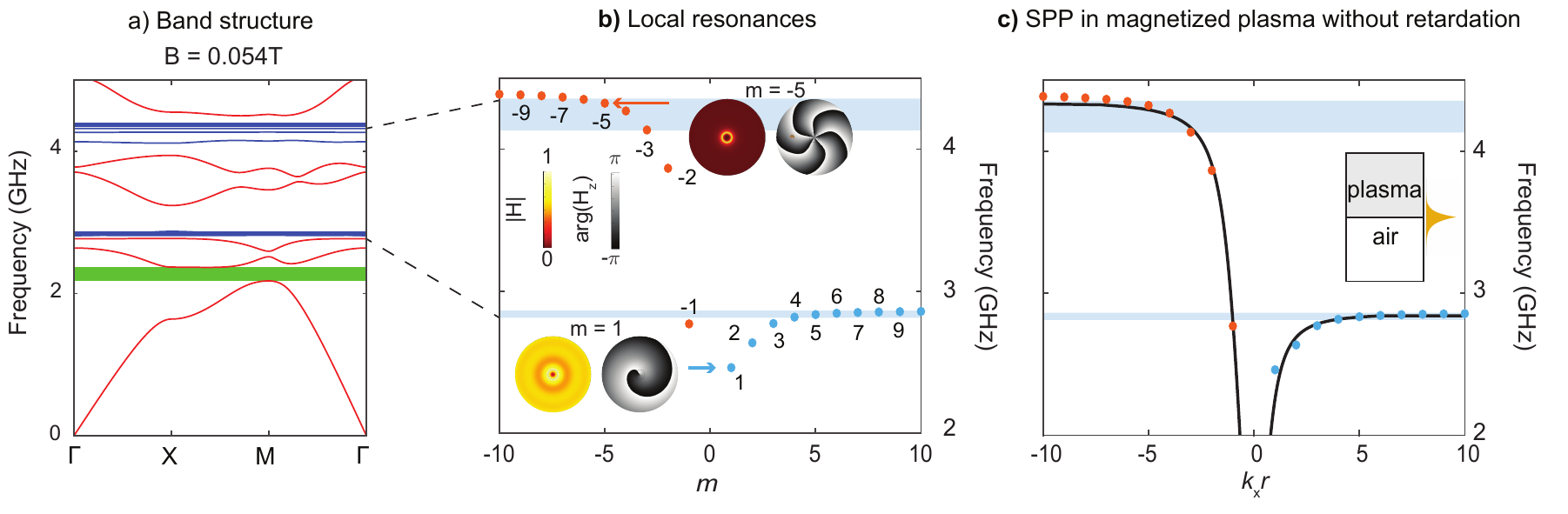}
    \caption{%
        \textbf{Band structure of magnetized-plasma photonic crystal featuring a Chern insulator gap.}
        \textbf{(a)} Under an external magnetic field of $B = 0.054\,\text{T}$, a full  energy gap is opened (green ribbon), featuring a non-zero Chern number. 
        Meanwhile, the flat bands split into two groups (blue ribbons). 
        \textbf{(b)}~%
        Local resonances, labeled by different azimuthal numbers $m$, are responsible for observed flat bands in \textbf{a}. 
        Two example mode profiles ($m=1$ and $m=-5$) are shown, both in  amplitude (hot color map) and phase (gray-scale). 
        \textbf{(c)}~The frequency of local resonances agrees well with the surface plasmon polariton (SPP) dispersion at the interface between  air and magnetized plasma.
        \label{fig:bandstructure_magnetized}
    }
\end{figure}

An example of the calculated band structures is shown in Fig.~\ref{fig:bandstructure_magnetized}a when the external magnetic field is set to be $B = 0.054\,\text{T}$. As time-reversal symmetry $T$ is broken, the $M$-point degeneracy is lifted, opening an 8\% full energy gap, from 2.18 to 2.37\,GHz (green ribbon). 
As the structure still maintains $C_{4z}$ symmetry, the Chern number $C$ of the first band is necessarily non-trivial\cite{PhysRevB.86.115112}, since 
\begin{equation}
\mathrm{e}^{\mathrm{i}\pi C} = C_{2z} (\Gamma) \times C_{2z} (M) = -1\optional,
\end{equation}
where $C_{2z} (\Gamma)=1$ since the phase of an electromagnetic wave is locked at zero frequency, and $C_{2z} (M)=-1$ because it originates from the time-reversal symmetry breaking of the degenerate modes shown in Fig.~\ref{fig:bandstructure_timereversal}b.
As a result, the first gap highlighted in green corresponds to a Chern insulator~\cite{PhysRevB.77.235125} and supports unidirectional transport channels, as shown next.  
Due to the magnetic field, the flat bands split into two regions (two blue ribbons), as explained in detail in the next section. 

\section{Flat bands from localized surface plasmon polariton resonances}
\label{sec:flat-bands}

In this section, we elucidate the origin of the flat bands observed above, from the viewpoint of localized surface plasmon polaritonic (SPP) resonances. We note that such flat band features are also commonly observed elsewhere, such as in metallic photonic crystals. The one unusual feature is related to the splitting of flat band regions under an external magnetic field (Fig.~\ref{fig:bandstructure_magnetized}a). 

It is more straightforward to understand the flat bands if we consider the local SPP resonances supported by a single plasma cylinder\cite{jin2016topological, jin2017infrared}. As the cylinder has full rotation symmetry, the SPP resonances can be labeled by different azimuthal numbers, $m$, corresponding to different angular momenta. 
A few interesting features can be observed in Fig.~\ref{fig:bandstructure_magnetized}b. 
First, the resonances with negative $m$ (orange dots, rotating counter-clockwise) are at higher frequencies than the resonances with positive $m$ (blue dots, clockwise). 
Second, the local SPP resonance frequency generally increases with $|m|$, approaching different bounds near the two ends: $(\sqrt{f_{\rm c}^2+2f_{\rm p}^2}+f_{\rm c})/2=4.4\,\text{GHz}$ when $m$ approaches $-\infty$ and $(\sqrt{f_{\rm c}^2+2f_{\rm p}^2}-f_{\rm c})/2=2.9\,\text{GHz}$ when $m$ approaches $+\infty$. 
Finally, the resonances with larger $|m|$s are better localized in space than the resonances with smaller $|m|$. Such a trend  can be well observed in the comparison between the mode profile of $m=-5$ (more localized) versus the model profile of $m=1$ (more extended, inset of Fig.~\ref{fig:bandstructure_magnetized}b). 
Taken together, at large $|m|$, the SPP resonances are well-localized near individual cylinders. 
Their modal overlaps are small, giving rise to little dispersion and thus the flat bands. 
Furthermore, the frequency of the flat bands (blue ribbons) also splits into two regions, approaching the two frequency bounds mentioned above. 

Here we note that the number of calculated flat bands using our numerical method increases with increased mesh density -- a common feature also observed in the literature\optional{~\cite{PhysRevB.50.16835}} -- although the frequency of the flat bands is always confined to the blue regions. Furthermore, the reliable azimuthal number of calculated modes is always limited by the numerical resolution. In our specific setting, modes with $|m|>10$ are no longer reliable. See Supplementary Information for more details. 

Finally, we verify the local SPP resonance frequencies in each plasma cylinder using the SPP dispersion along the interface between air and magnetized plasma, where a good agreement is found (Fig.~\ref{fig:bandstructure_magnetized}c). The momentum of each local SPP resonance is determined as $m/r$, where $m$ is the azimuthal number and $r$ is the radius of the cylinder. 

\section{Evolution of the chiral edge state dispersion with changing edge termination}
\label{sec:evolution}

While the existence of chiral edge states (CES) is guaranteed at the interface between a Chern insulator and a trivial insulator, their exact dispersion depends on the details of the interface. In this section, we continuously change the interface configuration and study how the CES dispersion evolves accordingly. Our finding suggests that CES dispersion essentially reflects that a localized plasma resonance emerges at the interface when the plasma is cut through, and the frequency of the antenna state decreases with the shrinking of the plasma region.

The CES dispersions are calculated from the interfaces between a Chern insulator super-cell and a pair of perfect magnetic conductors (PMC), as shown in Fig.~4. 
Each unit cell in the Chern insulator has the same design and parameters in Fig.~3. 
The lower interface (blue) is fixed, while the top interface (red) is continuously modified as the distance $d$ increases from $0$ to $a$. 
When $d/a=0$, a pair of CES are found in the super-cell dispersion, including one band at the top interface (red) and another at the bottom interface (blue). 
As $d/a$ increases, the CES on top (red) moves up in frequency until the PMC gets in contact with the plasma cell at $d/a = 0.25$. 
As $d/a$ keeps increasing, only the low-frequency portion of the CES continues to move up, while  the high-frequency part remains roughly unchanged until $d/a=0.5$ when the PMC cuts the plasma cell in half. 
For even larger $d/a$, the CES dispersion becomes flatter and flatter,  until when $d/a$ reaches $0.7$ and the CES splits into two bands: a CES band that traverses the topological gap, and a trivial band that is outside the topological gap. A second splitting occurs when $d/a=0.735$, and a second trivial band emerges.
As $d/a$ keeps increasing, both trivial bands keep going down in frequency, traveling through the first bulk continuum and eventually disappearing at zero frequency $f=0$ when $d/a=0.75$. See Supplementary Information for the case when $d/a=0.74996$.
Finally, when $d/a$ increases to $1$, the interface configuration becomes the same as $d/a=0$, and the CES dispersion goes back to the initial setting.  

Taken together, as $d/a$ increases from $0$ to $1$, both the first and the second band continua lose 1 mode, through the CES and the trivial edge state evolution. 
This can intuitively be understood as: when the super-cell loses a unit cell, each bulk continuum should also lose a mode, accordingly. 

\begin{figure}[htbp]
    \centering
    \includegraphics[width=\linewidth]{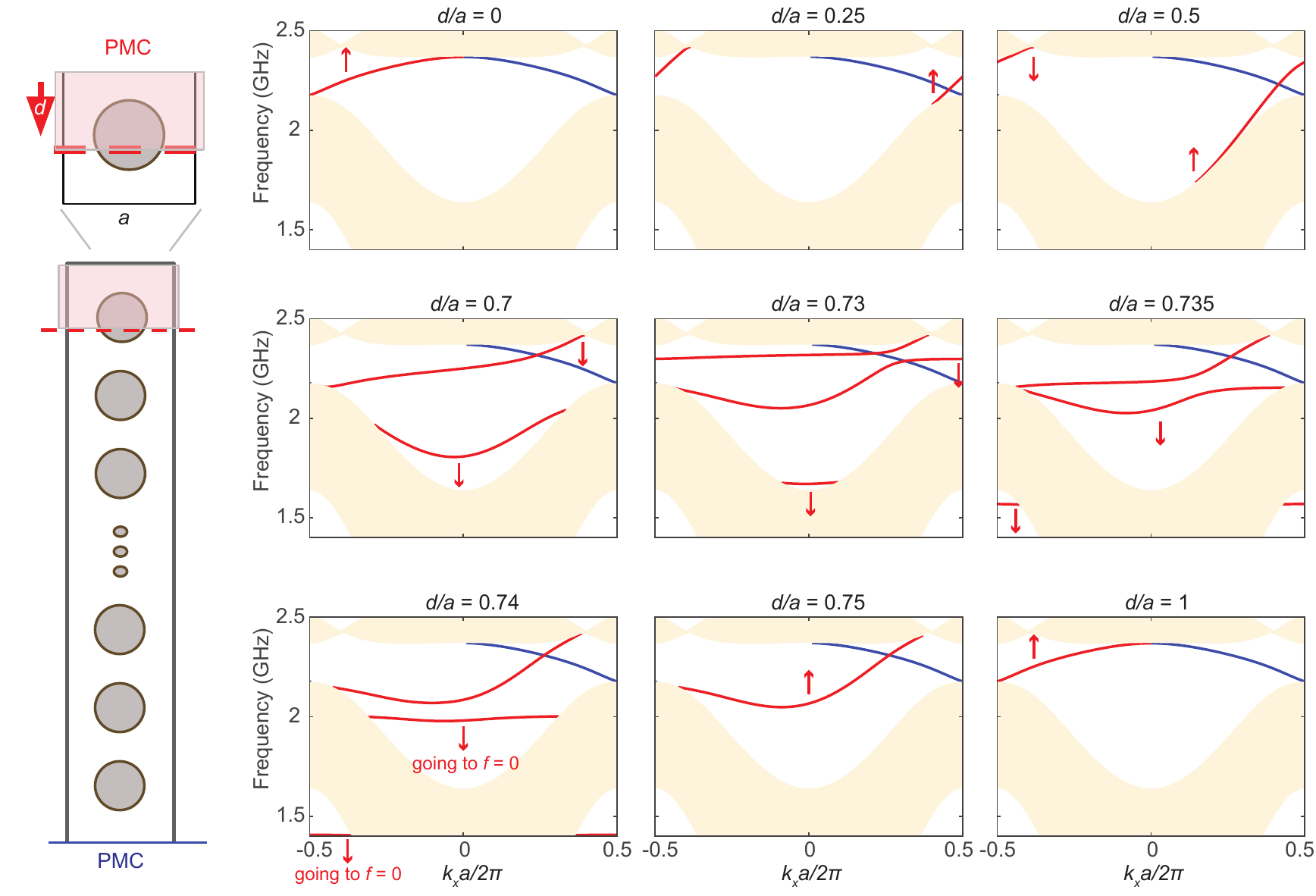}
    \caption{
        \textbf{Evolution of chiral edge states as the interface changes.}
        A super-cell of electromagnetic Chern insulator  (same design as in Fig. 3) is terminated by PMC. 
        The bottom interface (blue) is fixed, while the top interface (red) changes from $d/a = 0 $ to $d/a = 1$. 
        The corresponding chiral edge states (CES) dispersion under different configurations of $d/a$ is shown on the right.
        Red arrows indicate the evolution direction of the CES.
        \label{fig:edgestate_evolution}
        }
\end{figure}

\section{Discussion} 

Our simulation has a few limitations that should be noted. Firstly, the ions are assumed to be stationary and not moving, i.e., only electrons are allowed to move under the external fields. Introducing ions' motion would lead to an effective mass and lower plasma frequency. Secondly, our permittivity model ignores non-local effects, which results in a frequency gap between surface plasmon polaritons traveling in opposite directions\cite{HassaniGangaraj:19,PhysRevB.55.7427}. Thirdly, we have neglected the inhomogeneous broadening effect in the plasma, and the bandwidth of our permittivity is contributed solely by electron damping. Introducing the inhomogeneous broadening effect would further broaden the energy bands and reduce the effective size of the band gaps.

Despite these limitations, we believe that our proposal is feasible to demonstrate in an experiment. The required magnetic field of 0.05\,T can be achieved, even over large areas, via commercial electromagnets or permanent magnets. Meanwhile, the required carrier number density of $2.8\times 10^{11}\,\text{cm}^{-3}$ and the pressure of 0.05\,Torr are also within the typical range in experiments.

\begin{backmatter}
\bmsection{Funding}
Air Force Office of Scientific Research (FA9550-21-1-0299); Villum Fonden (42106).

\bmsection{Acknowledgments}
A.V.K. acknowledges fruitful discussions with M.G. Silveirinha.

\bmsection{Disclosures}
The authors declare no conflicts of interest.

\bmsection{Data availability} Data underlying the results presented in this paper are not publicly available at this time but may be obtained from the authors upon reasonable request.

\bmsection{Supplemental document}
See Supplement 1 for supporting content. 

\end{backmatter}


\bibliography{references}






\end{document}

%% file: misc-utils.tex
\usepackage{textcomp} 
\usepackage{xifthen}
\usepackage{etoolbox}
\newboolean{togglecomments}
\newboolean{toggletodos}
\newboolean{togglechanges}

\newcommand{\textblacksquare}{$\blacksquare$}
\newcommand{\todo}[1]{\ifbool{toggletodos}%
	{\textcolor{green!60!black}{\small\textsf{{}\textsuperscript{\textsc{\textsf{todo}}}}[\ignorespaces#1]}} 
	{}}     
\newcommand{\comment}[2]{\ifbool{togglecomments}%
		{\textcolor{blue!70!black}{\small\sf\textsuperscript{\textsc{\textsf{\ignorespaces#1}}}[\ignorespaces#2]}} 
		{}}     
\newcommand{\swap}[2]{\ifbool{togglechanges}
	{\ignorespaces#2}  
	{\textcolor{red!70!black}{[\ignorespaces#1]}%
         \textrightarrow{}%
         \textcolor{green!50!black}{[\ignorespaces#2]}
        }}
\newcommand{\remove}[1]{\ifbool{togglechanges}
	{}    
	{\textcolor{red!70!black}{\ignorespaces#1}}}
\newcommand{\inset}[1]{\ifbool{togglechanges}
	{\ignorespaces#1}  
	{\textcolor{green!50!black}{\ignorespaces#1}}}

\newcommand{\citeremind}[1]{%
	[\textcolor{blue!75!black!80!yellow}{\textblacksquare%
		\ifthenelse{\isempty{#1}}{}{\textsuperscript{\tiny\textsf{\ignorespaces#1}}}%
	}]}

\newcommand{\optional}[1]{\textcolor{orange!70!gray}{\ignorespaces#1}}